\documentclass[review]{elsarticle}
\usepackage{graphicx}
\usepackage{caption}
\usepackage{subcaption}
\usepackage{hyperref}
\usepackage{listings}
\usepackage{xcolor}
\journal{Journal of Computational Physics}

\begin{document}

\begin{frontmatter}

\title{The swept rule for breaking the latency barrier in time advancing PDEs}


\author[address]{Maitham Alhubail\corref{mycorrespondingauthor}}
\cortext[mycorrespondingauthor]{Corresponding author}
\ead{hubailmm@mit.edu}

\author[address]{Qiqi Wang}
\ead{qiqi@mit.edu}
\ead[url]{www.engineer-chaos.blogspot.com}

\address[address]{Department of Aeronautics and Astronautics, MIT, 77 Massachusetts Ave., Cambridge, Massachusetts 02139, USA}

\begin{abstract}
This article investigates the swept rule of space-time domain decomposition, an idea to break the latency barrier via communicating less often when explicitly solving time-dependent PDEs. The swept rule decomposes space and time among computing nodes in ways that exploit the domains of influence and the domain of dependency, making it possible to communicate once per many timesteps without redundant computation.  The article presents simple theoretical analysis to the performance of the swept rule which then was shown to be accurate by conducting numerical experiments.
\end{abstract}


\end{frontmatter}


\section{Introduction}

Today, extreme-scale computer clusters can solve PDEs using over 1,000,000 cores\cite{million}. Exa-scale computing promises to enable 1,000,000,000-core simulations\cite{exascale1,exascale2}. From an application perspective and compared to desktop-based tools, using such computing power resembles upgrading from a slide rule to a desktop computer\cite[Chapter~6]{sliderule}.\newline

There is strong demand to use the increasingly parallel computing power to accelerate solutions of unsteady PDEs. In designing rocket launch vehicles\cite{launch}, for example, unsteady flow equations are solved to reduce unsteady aerodynamic loads on the vehicle structure. In designing jet engine combustors\cite{lescomb}, unsteady reacting flow equations are solved to help reduce emission and mitigate dangerous vibrations. In designing gas turbine blades\cite{blades1,blades2}, unsteady aero and thermodynamic equations are solved to design cooling mechanisms for increased component life and overall engine efficiency. In all these applications, the unsteady equations need to be solved long enough to reproduce the chaotic, multiscale fluid dynamics. That requires at least hundreds of thousands, often millions of time steps. Advancing this many time steps with today's technology takes days, weeks, sometime months, running on as many parallel processors as these simulations can effectively use. The time-to-solution of these simulations often becomes the bottleneck of new product development and technological innovation. Engineers desire to further scale these simulations, to run them faster.\newline

What is preventing these simulations from scaling? If the total amount of computation required to solve a PDE is fixed, and we divide the computation among twice as many computing nodes, shouldn't each node work half as much, and the PDE solved in half the time? This would be true if each node does its own work without communicating with others. In solving PDEs, however, nodes communicate to each other frequently. Each communication takes at least a few microseconds, and on a common network, tens of microseconds. This minimum communication time, regardless of how much information is communicated, is called the network latency. If network latency exceeds the computing time between consecutive communications, reducing the computation of each node does not accelerate the simulation. This barrier to scaling, called the latency barrier, is necessarily encountered as a PDE-based simulation is deployed to more nodes.\newline

This latency barrier impacts simulations with many time steps, such as high fidelity, unsteady computational fluid dynamics simulations.  Among applications, the latency barrier particularly hurts those in which fast turn around time is critical, such as design optimization. Network latency improves slowly over time compared to the Moore's law. On average, network bandwidth doubles every two years; but over the same amount of time, latency improves on average by no more than 20 to 40 percent\cite{latency}. Cloud computing has the potential to enable extreme scale computing without dedicated supercomputers\cite{cloud}. The latency in cloud computing, however, is typically worse than dedicated clusters, making it more important to break the latency barrier\cite{cloudlatency}.\newline

This article investigates the swept rule, an idea to break the latency barrier via communicating less often. The swept rule decomposes space and time among computing nodes in ways that exploit the domains of influence and the domain of dependency, making it possible to communicate once per many time steps. The resulting algorithm enables simulations to be solved significantly faster than what is possible with spatial domain decomposition schemes typically found in today's PDE solvers.\newline

The same goal is shared by parallel-in-time methods\cite{parareal,mgrit,pfasst,50years,realtime,adjoint}. These methods first estimate the solution using cheap but inaccurate time integrators, then iteratively correct the solution with an expensive and accurate time integrator.  Parareal is one of the most studied parallel in time algorithms and was proposed by by Lions, Maday, Turinici in 2001\cite{parareal}.  One advantage of the Parareal method is that it allows the use of existing classical time-stepping routines and yet gain parallelism.  It is also important to mention that, due to Parareal’s popularity, a Parareal-like algorithm has been extended to chaotic dynamical systems\cite{pararealextended}.  Another famous parallel in time algorithm is the Multigrid Reduction in Time (MGRIT).  One major difference between the MGRIT and Parareal is the use of multilevel time integrators\cite{mgrit}.
The third famous parallel in time method is the Parallel Full Approximation Scheme in Space and Time (PFASST).  In away, PFASST is similar to Parareal except that it uses advance iterative time stepping and full approximation schemes that allow for making the cheap inaccurate time integrators as cheap as possible\cite{pfasst}.  These three parallel in time methods share being iterative and involving a coarse solve.\newline

The method presented in this paper, technically differ from most parallel-in-time methods, in that it is not iterative, and does not involve a coarse solver.  Parallel-in-time algorithms iterate over the time domain.  These methods work best with PDE systems that do not have extreme non-linearity in their dynamics.  For example, parallel-in-time algorithms are less efficient when solving chaotic systems.
The Swept Rule method advances in time and does not iterate over time.  Therefore, it works equally well to different dynamical systems and different schemes.  It perhaps has more commonality with Communication Avoiding (CA) algorithms\cite{ca1,ca2}.  When it comes to solving a PDEs, CA algorithms are found to contribute and focus more on communication avoidance for the basic operations of linear algebra.  Those algorithms include but are not limited to LU, QR, Matrix-Matrix multiplication and Matrix-Vector multiplication.  CA algorithms cover a wide spectrum of techniques to achieve their goal in communicating less frequently starting from the different levels of memory hierarchy, to CPU-GPU communication and up to node to node communications.  Cache oblivious programming and computation domain overlapping are just two examples of the techniques that CA algorithms use to minimize or avoid communication latency while performing a specific linear algebra task\cite{ca1,ca2}.\newline

When it comes to our objective, that is breaking the latency barrier when explicitly solving PDEs, we see that swept decomposition is different from CA algorithms in that it does not involve any overhead of overlapping parts of the computation domain among processors, which is the case in some CA algorithms.  In fact, the way swept is designed and implemented makes the effort of developing an explicit numerical PDE solver based on the swept decomposition similar to that effort involved in developing a classic-straight decomposition based solver.
\section{Space-time decomposition of a PDE solver}

We restrict our attention to the following case. A PDE is discretized with a finite difference, finite volume, or finite element scheme. These discretization schemes generate a graph in the spatial domain. Each point in this graph represent a grid point in finite difference, a control volume in finite volume, or an element in finite element. An edge between two points exists if they are neighboring grids, control volumes or elements, as defined by the spatial discretization.\newline

\begin{figure}
\centering
\includegraphics[width=0.7\linewidth]{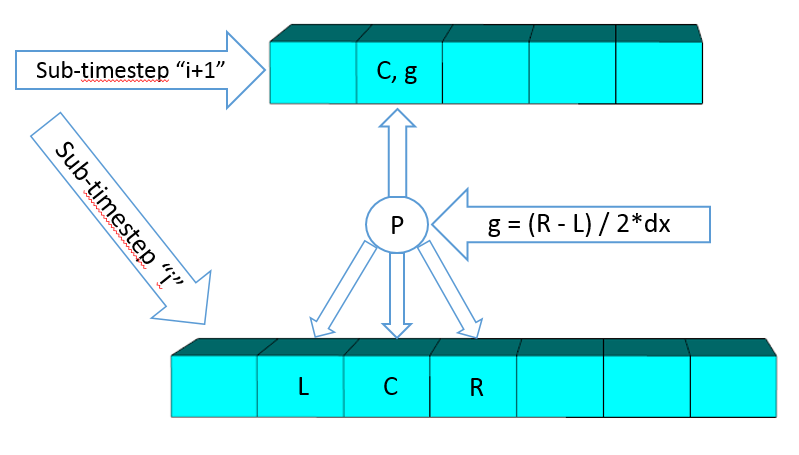}
\caption{Illustration of Input, Output forward and variable calculations}
\end{figure}

To analyze and break the latency barrier, we attempt to decompose the solution procedure into elementary steps that require communication between computing nodes. We call such an elementary step a sub-timestep. Each sub-timestep has a set of input variables and a set of output variables at each spatial point. From the input variables at each point and its neighbor points, the sub-timestep computes its output variables at that point and forwards any variables from the previous sub-timestep to the next sub-timestep if needed.  Such variable forward thereby allows decomposition of a complex timestep into sub-timesteps, eliminating the need for neighbors of neighbors within a single sub-timestep.  Figure 1 clarifies the above explanation.  In the figure, the point ``P" forwards the input of the sub-timestep below it to the sub-timestep above it as the first input.  ``P" also calculates the gradient using its left and right neighbors and sends that to the sub-timestep above it as the second input.\newline

For example, finite difference gradient computation on a compact stencil is such a ``substep". The inputs are the value of field variables; the outputs are the gradient of these variables. The following code exemplifies such a sub-timestep on a one-dimensional uniform grid. Here ``p" denotes a spatial point; ``p.Linputs" and ``p.Rinputs" denote its left and right neighbors; ``p.inputs[i]" and ``p.outputs[i]" denote the ``i" th input and output variable at the spatial point ``p".  Moreover, ``p.Linputs[i]" and ``p.Rinputs[i]" denote the ``i" th inputs of the left and right neighbors respectively.

\begin{lstlisting}
p.output[0] = p.inputs[0]; 
p.output[1] = (p.Rinputs[0] - p.Linputs[0]) / 2 * dx;
\end{lstlisting}

This sub-timestep has one input and two output variables. The first output variable is a carbon copy of the input; the second output variable is the finite difference derivative of the input.\newline

Another example of a ``substep" is flux computation and accumulation in finite volume.  The inputs are the conservative variables and their derivatives; the outputs are the divergence of fluxes, or time derivatives of the conservative variables. The following code illustrates a substep that applies Godunov's finite volume scheme to a scalar conservation law, and uses forward Euler for time integration.

\begin{lstlisting}
fluxL = riemannSolver(p.Linputs[0] , p.inputs[0]);
fluxR = riemannSolver(p.inputs[0]  , p.Rinputs[0]);
p.outputs[0] = p.inputs[0] + dt * (fluxR - fluxL) / dx;
\end{lstlisting}

The input of this sub-timestep is the conservative variable at the $i$th time step, and the output is the conservative variable at the $i+1$st time step.\newline

If the forward Euler time integration is replaced by a multi-stage Runge-Kutta scheme, then each timestep require multiple sub-timesteps. In a two-stage Runge Kutta scheme, also known as the midpoint method, the first sub-timesteps can be

\begin{lstlisting}
fluxL = riemannSolver(p.Linputs[0] , p.inputs[0]);
fluxR = riemannSolver(p.inputs[0]  , p.Rinputs[0]);
p.outputs(0) = p.inputs(0);
p.outputs(1) = p.inputs(0) + 0.5 * dt * (fluxR - fluxL) / dx;
\end{lstlisting}

It has a single input and two outputs, the first being a carbon copy of the input, and the second being the solution at the midpoint. These outputs become the inputs of the the second sub-timestep, whose single output is the solution at the next timestep:

\begin{lstlisting}
fluxL = riemannSolver(p.Linputs[1] , p.inputs[1]);
fluxR = riemannSolver(p.inputs[1]  , p.Rinputs[1]);
outputs[0] = p.inputs[0] + dt * (fluxR - fluxL) / dx;
\end{lstlisting}

Every sub-timestep must use the same inputs as the outputs of the previous one.  This means some sub-timesteps must ``forward" some variables not involved in computation, by setting some of its outputs to the values of their corresponding inputs.\newline

If we replace Godonov's scheme with a second order finite volume scheme, and use the forward Euler time discretization, then two sub-timesteps is required. The first sub-timestep can be similar to the finite difference derivative computation shown above, whose two outputs are the conservative variable itself and its derivative. The second sub-timestep, whose inputs are the outputs of the previous sub-timestep, can be encoded as following,

\begin{lstlisting}
uL_minus = reconstructWithLimitor(p.Linputs[0],
                  p.inputs[0],p.Linputs[1], dx);
uL_plus  = reconstructWithLimitor(p.inputs[0],
                  p.Linputs[0], -p.inputs[1], dx);
fluxL    = riemannSolver(uL_minus, uL_plus);

uR_minus = reconstructWithLimitor(p.inputs[0], 
                  p.Rinputs[0],    p.inputs[1], dx);
uR_plus = reconstructWithLimitor(p.Rinputs[0], 
                 p.inputs[0], -p.Rinputs[1], dx);
                 
fluxR = riemannSolver(uR_minus, uR_plus);
p.outputs(0) = p.inputs[0] + dt * (fluxR - fluxL) / dx;
\end{lstlisting}

Combining the second order finite volume scheme with a two-stage Runge-Kutta would result in two sub-timesteps per Runge-Kutta stage, thus four sub-timesteps per timestep. Solving a system of conservation laws, e.g., Euler equation, would require more inputs and outputs for each sub-timestep. But the total number of sub-timesteps would be no different from a scalar conservation law discretized with a similar scheme.\newline

When such decomposition is possible, each sub-timestep can be viewed as an elementary operation in solving a PDE. The number of sub-timesteps involved in a calculation determines its domain of dependence and the domain of influence. Consider the outputs of the $m$th sub-timestep on the $i$th spatial point (grid point, cell or element). Its domain of dependence covers only its immediate neighbor points over one sub-timestep, i.e., among the inputs of the $m$th sub-timestep (or equivalently, the outputs of the $m-1$st sub-timestep). Over two sub-timesteps, the domain of dependence expands to the neighbors of neighbors. Over $n$ sub-timesteps, it grows to all spatial points that connect to point $i$ through $n$ or fewer edges. 

\begin{figure}
\centering
\includegraphics[width=0.7\linewidth]{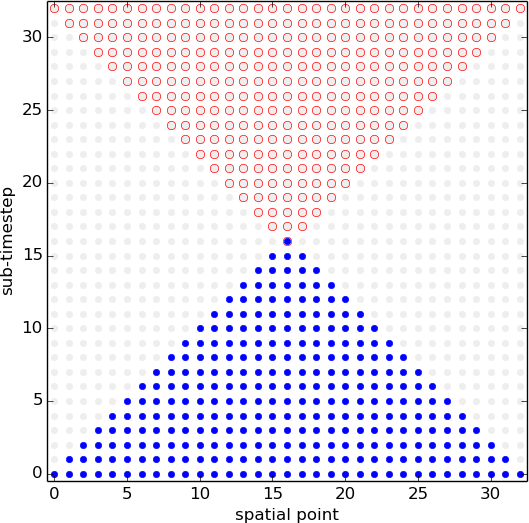}
\caption{The domains of dependence and influence in a one-dimensional spatial domain}
\end{figure}

Figure 2 shows the domains of dependence in a one-dimensional spatial domain, as well as the domain of influence. The swept boundaries of these domains in the space-time diagram motivates the ``swept" rule for solving PDEs. This rule decomposes along such space-time boundaries, and assigns the decomposed chunks of space-time among processors, thereby avoiding frequent latency-incurring communications.

\section{The swept rule in 1D}

\begin{figure}
	\centering
	\begin{subfigure}[b]{0.45\textwidth}
	\includegraphics[width=1.0\textwidth]{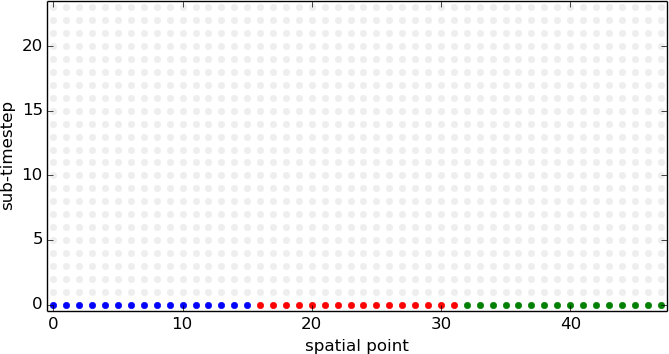}
	\caption{Setting initial condition}
	\end{subfigure}
	\quad
	\begin{subfigure}[b]{0.45\textwidth}
	\includegraphics[width=1.0\textwidth]{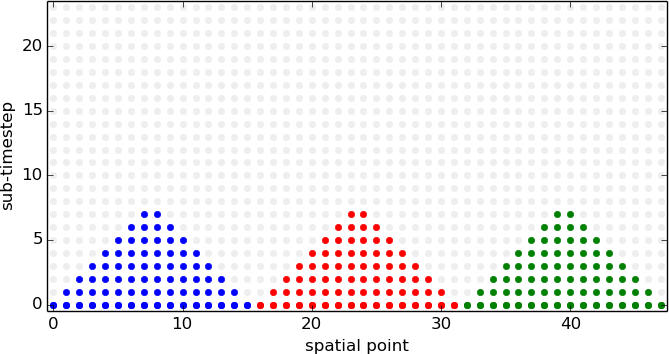}
	\caption{1st local computation}
	\end{subfigure}

	\begin{subfigure}[b]{0.45\textwidth}
	\includegraphics[width=\textwidth]{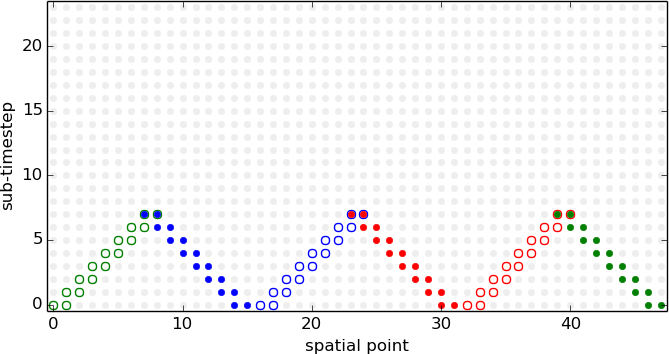}
	\caption{1st communication}
	\end{subfigure}
	\quad
	\begin{subfigure}[b]{0.45\textwidth}
	\includegraphics[width=\textwidth]{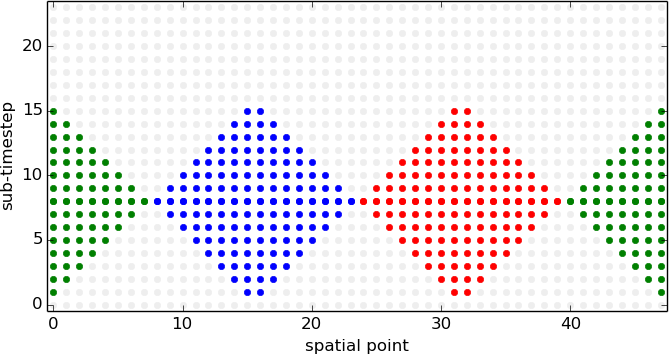}
	\caption{2nd local computation}
	\end{subfigure}
	
	\begin{subfigure}[b]{0.45\textwidth}
	\includegraphics[width=\textwidth]{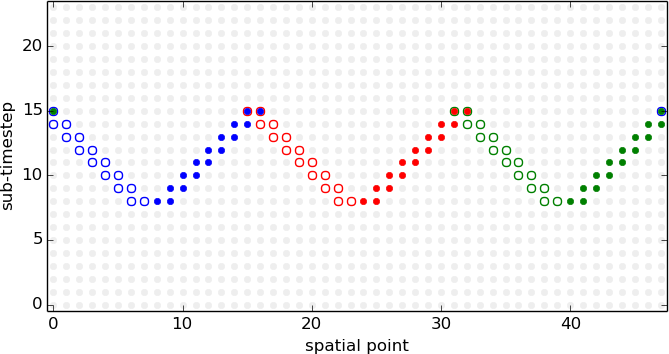}
	\caption{2nd communication}
	\end{subfigure}
	\quad
	\begin{subfigure}[b]{0.45\textwidth}
	\includegraphics[width=\textwidth]{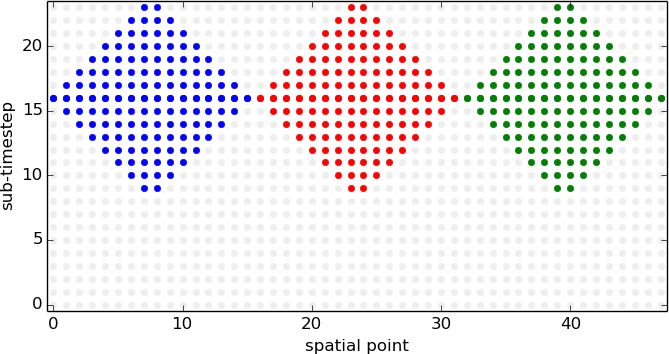}
	\caption{3rd local computation}
	\end{subfigure}
	
	\begin{subfigure}[b]{0.45\textwidth}
	\includegraphics[width=\textwidth]{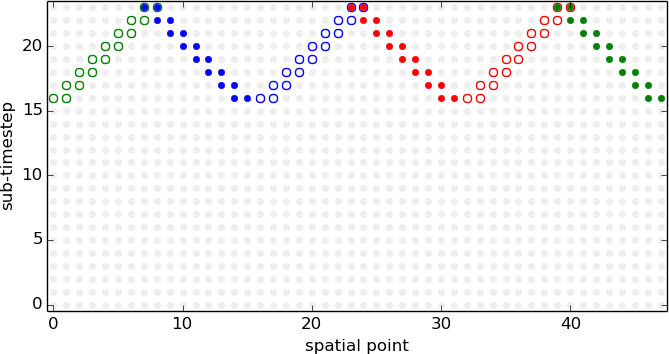}
	\caption{3rd communication}
	\end{subfigure}
	\quad
	\begin{subfigure}[b]{0.45\textwidth}
	\includegraphics[width=\textwidth]{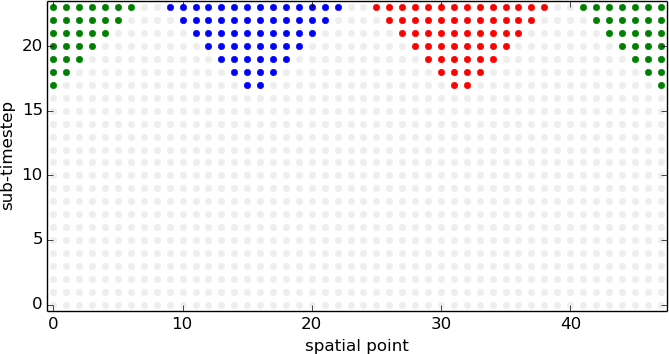}
	\caption{4th local computation}
	\end{subfigure}
	
    \caption{The computing and communication stages for 1D Swept rule}
\end{figure}

For illustration, we consider a one-dimensional spatial domain with periodic boundary condition. It is discretized into a series of spatial points, each representing a grid point in finite difference, a control volume in finite volume, or an element in finite element. Each point has two immediate neighbors, a left and a right one. When the PDE is initialized, the spatial points are decomposed uniformly among computing nodes, each node owning an even number of points. Each node starts by initializing the PDE on the points it owns.\newline

This initial decomposition is illustrated in Figure 3(a). In these figures, a circle at horizontal coordinate $i$ and vertical coordinate $t$ represents the input variables of the $t$th sub-timestep at the $i$th spatial point, which for $t=0$ are set by the initial condition of the PDE, and for $t>0$ are the output variables of the $t-1$st sub-timestep. At $t=0$, the input variables of the first sub-timestep is colored by the computing node storing it. The variables at later sub-timesteps are greyed, indicating that they are not yet computed.\newline

After setting the inputs of the first sub-timestep, we start the first local computation. During this stage, each parallel computing node computes all it can compute without communicating with other nodes. It computes all variables of which, at initialization, it owns the domain of dependence. These variables are colored accordingly in Figure 3(b). They include the output of the first sub-timestep on all but the two boundary points, the output of the second sub-timestep on all but four points, and the output of subsequent sub-timesteps on a shrinking subset of points. The stage completes when the output variables of a sub-timestep are computed at only two points, and nothing more can be computed without communicating with neighboring nodes.\newline

Then, for the first time, the computing nodes communicate. Each node packs a subset of its computed variables and sends them to its neighboring node on the left. The transferred variables include two left-most outputs of every sub-timestep, forming a swept leading-edge of the space-time domain covered by the previous computing stage. In addition to sending the left leading-edge, each node keeps in its memory another swept leading-edge on the right side of the space-time domain. This right leading edge, represented by solid circles in Figure 3(c), combines with the left leading edge received from its right neighboring node, represented by open circles of the same color, to provide the variables needed for the next local computing stage.\newline

The second local computation, like the first one, proceeds independently on each computing node. Each node starts with a V-shape in the space-time domain, formed by a pair of swept leading-edges, the left inherited from the first local computation, and the right received during the first communication. As in every local computing stage, each node computes all it can compute without communicating with other nodes. It computes all variables, lying in a space-time diamond illustrated in Figure 3(d), whose domains of dependence is covered by the pair of swept leading-edges forming the V. As in other local computing stage, this stage completes when the output variables of a sub-timestep are computed at only two points, and nothing more can be computed without communicating again with neighboring nodes.\newline

The second communication is just like the first one, except that the right swept leading-edge is sent to the right neighboring node, and the left swept leading edge is kept in its own memory. In subsequent communication steps, even-numbered stages send the right leading-edge towards the right neighboring node, and odd-numbered rule stages send the left leading-edge towards the left neighboring node. This rule ensures each node to alternate between two sets of spatial points during odd and even numbered computing stages, limiting each node to two sets of spatial points. The result of the second communication stage is illustrated by Figure 3(e), in which the solid circles represent variables each node keeps from the previous computing stage, and the open circles of the same color represent variables the same node receives from its neighboring node.\newline

The subsequent computing stages and communication stages, illustrated in Figures 3(f) to 3(h), are similar to the previous ones. Each computing stage compute all each node can compute without communication; each communication stage transfers a swept leading-edge over each pair of neighboring nodes, enabling the next computing stage.
\section{A simplified performance analysis of the swept rule}

To qualitatively understand the performance of the swept rule, we perform a simplified analysis, by making the following two assumptions:
\begin{enumerate}[1. ]
\item Communication between computing nodes takes time $\tau$, regardless of how much data is communicated.
\item Each sub-timestep on each spatial point takes time $s$ to compute.
\end{enumerate}

Let $N$ be the total number of spatial points, and $p$ be the number of computing nodes. Then the number of spatial points per node is $n=N/p$. A cycle of the swept rule, advancing the PDE for $n$ sub-timesteps, consists of two computing stages and two communication stages. The two computing stages perform a total of $n^2$ calculations per node, each applying a sub-timestep to a spatial point. The computing stages therefore take $n^2 s$ time, according to our simplifying assumption. The two communication stages take $2\tau$ time.  The entire cycle then takes $n^2 s + 2\tau$ time. Divided among the n sub-timesteps during the cycle, the amount of computing time per sub-timestep is:
\begin{equation}
n s + 2 \frac{\tau}{n}
\end{equation}

By increasing or decreasing how many nodes we use, we can easily change $n$. The other two variables, $\tau$ and $s$, are set by the hardware and the discretization of the PDE; they are harder to change. To understand how fast the swept rule is, we need to know the typical values of $\tau$ and $s$.\newline

\begin{table}
\begin{center}
\begin{tabular}{|l|l|}\hline

Interconnect                   & Typical latency ($\tau$) \\\hline
Amazon EC2 cloud	           & 150 $\mu s$               \\\hline
Typical Gigabit Ethernet	   & 50  $\mu s$               \\\hline
Fast 100-Gigabit Ethernet	   & 5   $\mu s$               \\\hline
Mellanox 56Gb/s FDR InfiniBand & .7  $\mu s$               \\\hline

\end{tabular}
\end{center}
\caption{The range of latency $\tau$ commonly encountered today.}
\end{table}

Table 1 attempts to cover the range of latency $\tau$ one may encounter today.  The latency can change over three orders of magnitude, from the fastest Infiniband to a cloud computing environment not designed for PDEs.
\newline
The range of $s$ is even wider; it can span over eight orders of magnitude. $s$ depends both on the computing power of each node and on the complexity of each sub-timestep. If one sub-timestep on one spatial point takes $f$ floating point operations (FLOP) to process, then processing an array of them with a node capable of $F$ floating point operations per second (FLOPS) takes $s = f/F$ seconds per step-point. Running a cheap discretization on a powerful computing node leads to small $f$ and large $F$, therefore a small $s$; running an expensive discretization on a light node leads to large $f$ and small $F$, therefore a large $s$.\newline

Table 2 attempts to estimate the range of $s$ by covering the typical $f$ for solving PDEs in a single spatial dimension, as well as the highest and lowest $F$ on a modern computing node. Consider $f$ of the heat equation, discretized with finite difference. Each sub-timestep takes only 3 FLOPs. A sub-timestep of a nonlinear system of equations, discretized with high order finite element, may take thousands of FLOPs. The highest $F$ is achieved today by GPU nodes. 11.5 TeraFLOPS has been achieved by the AMD Radeon R9 295X2 graphics card, as well as by combining two AMD Radeon R9 290X graphics cards in a single node. The Summit supercomputer, expected to be delivered to Oak Ridge Leadership Computing Facility (OLCF) in 2017, uses a similar architecture to achieve 40 TeraFLOPS per node. Using older CPU architecture is slower. Particularly slow is equivalence of using a single thread per node, e.g., in flat-MPI-style parallel programming. On the first generation Intel Core i3/i5/i7 architecture, codenamed Nehalem, a single thread is capable of about 10 GFLOPS. These different node architectures, running different PDE discretization schemes, yield orders of magnitude different values of $s$.

\begin{table}
\begin{center}
\begin{tabular}{|c|c|c|}\hline

Computing node -- FLOPS & FLOP per step-point & Computing time\cr & & per step-point ($s$) \\\hline
Single thread Intel Nehalem & & \cr 10 GFLOPS & 4000 (FE system) & 400 $ns$ \\\hline
Single thread Intel Nehalem & & \cr 10 GFLOPS & 200 (FV system)  & 20 $ns$  \\\hline
Single thread Intel Nehalem & & \cr 10 GFLOPS & 3 (FD scalar)    & 0.3 $ns$ \\\hline
Oak Ridge 2017 Summit node & & \cr  40 TFLOPS & 4000 (FE system) & 100 $ps$ \\\hline
Oak Ridge 2017 Summit node & & \cr  40 TFLOPS & 200 (FV system)  & 5 $ps$   \\\hline
Oak Ridge 2017 Summit node & & \cr  40 TFLOPS & 3 (FD scalar)    & 75 $fs$  \\\hline

\end{tabular}
\end{center}
\caption{The typical time it takes to compute one sub-timestep on a single spatial point for solving PDEs in a single spatial dimension}
\end{table}

With these values of $\tau$ and $s$, the plot in Figure 4 shows, according to Equation (1), how fast the diamond scheme runs as a function of $n$, the spatial points per node. The up-sloping, dashed and dash-dot lines represent the $n s$ term in Equation (1) for different values of $s$, and the down-sloping, solid lines represent the $\tau/n$ term for different values of $\tau$. For each combination of $s$ and $\tau$, the total time per sub-timestep as a function of $n$, plotted as thin black curves in Figure 4, can be found by summing the corresponding up-sloping and down-sloping lines.

\begin{figure}
\centering
\includegraphics[width=1.0\linewidth]{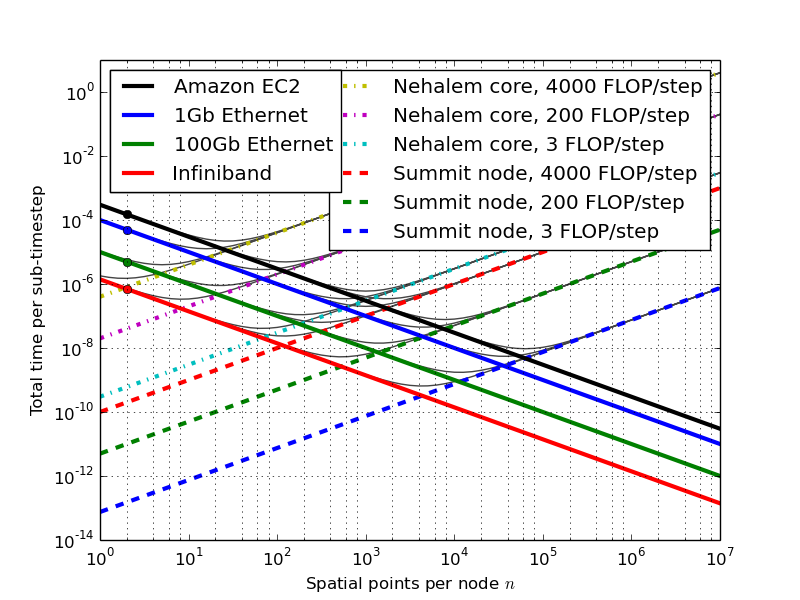}
\caption{Analyzing the diamond performance using communication latency and CPU FLOP/step}
\end{figure}

The total time per sub-timestep can be minimized by choosing $n$. This minimizing $n$ can be found, for each $\tau$ and $s$, at the intersection of the corresponding upsloping and downsloping lines in Figure 4. It has a mathematical expression $n^* =\sqrt{2\tau/s} $. Is largest for low $s$ and high $\tau$. $n^*$ can be as large as tens of thousands when advancing a simple PDE with a cheap discretization on most powerful GPU instances in a cloud computing environment. $n^*$ is smallest for high $s$ and low $\tau$. It can be as small as 1 or 2 when advancing a complex PDE with an expensive discretization in a older-CPU-based flat-MPI computing environment. These optimal values of $n$ represent the limit of scaling. Above this optimum, decreasing $n$ by scaling to more nodes would decrease the total time per sub-timestep, accelerating the simulation. But decreasing $n$ beyond the optimum by scaling to even more nodes would not accelerate, but slow down the simulation.\newline

At the optimal $n$, the minimum total time per sub-timestep also depends on $\tau$ and $s$. It has a mathematical expression $t^* = \sqrt{8\tau s}$. The fastest integration, unsurprisingly, is achieved for the lowest $s$ and lowest $\tau$. What is surprising is how fast it can be. The swept rule is theoretically capable of about 1 nanosecond per step, or almost a billion steps per second, if it uses the fastest Infiniband and the most powerful GPU nodes to efficiently advance the simplest PDE. This is about three orders of magnitude faster than what can be achieved with a method that requires communication every sub-timestep. To achieve this theoretical speed limit, it is necessary to not only minimize the latency $\tau$, but also minimize $s$ by fully utilizing the computational throughput of the most powerful computing nodes.\newline

At the optimal $n$, the swept rule almost always breaks the latency barrier. A method that requires communication every sub-timestep takes at least $\tau$ per sub-timestep. This limit is the latency barrier. The swept rule, according to our simplified model, takes $t^* = \sqrt{8\tau s}$ per sub-timestep. It breaks the latency barrier whenever $\tau > 8s$, i.e., when the network latency exceeds the time it take for a computing node to advance one sub-timestep on 8 spatial points. In that case, it breaks the latency barrier by a factor of $\sqrt{\tau/8s}$. This ratio is largest for small $s$ and large $\tau$, i.e., when the discretization is cheap, each computing node is powerful, and the latency between nodes is high. For example, if Amazon EC2 upgrades their GPU instances to the most powerful available, using them to advance the simplest finite difference equation would break the latency barrier by a factor of about 13,000.\newline

The swept rule has the potential to achieve 1 nanosecond per sub-timestep and breaking the latency barrier by a factor of 13,000. To get a realistic number for practical applications, we may assume that the discretization requires 100 FLOP per sub-timestep, and with good software engineering, a third of the computing power in a powerful node can be effectively utilized. Then with the fastest Infiniband, the swept rule can hope to achieve 10 nanosecond per sub-timestep, and with a cloud-computing-like latency, break the latency barrier by a factor of 1,300.

\section{Interface and implementation of the swept scheme}

The swept rule may seem challenging to implement, but it does not have to be. If a PDE solver can be decomposed into sub-timesteps, then the code that executes a sub-timestep on a spatial point can separate from the code that orders these executions and communicates with other computational nodes. The former code, which operates locally in space and time, implements the numerical scheme independent of the computer architecture. The later, global code, which decides when and where the local operations are executed, is optimized for computer architecture but is blind to what numerical scheme is used or what different equations is solved. The local and global codes share only an interface.\newline

As of our current implementation of the swept rule, our interface consists of two simple functions.  The first function will be called once per special point to set its initial data.  The input variables to the initialization function are the global special point index and a special point structure.  The following psedo-code exemplifies the initialization function interface
\begin{lstlisting}
Init(pointIndex i, spacialPoint p)
{
	p.inputs[0] = variable 0 initial value
	p.inputs[1] = variable 1 initial value
	p.inputs[n] = variable n initial value
};
\end{lstlisting}

The second function in our interface is where the PDE solve actually happens.  The input variables to the timestepping function are the index of which sub-timestep to be executed and a spacial point structure.  The following psedo-code exemplifies the timestepping function of our interface.

\begin{lstlisting}
timestep(substepIndex i, spacialPoint p)
{
#Based on the value of i, perform the proper operation in "p"
};
\end{lstlisting}

It is the responsibility of the swept rule scheme implementation developer to call the timestepping function with the proper sub-timestep index and spacial point.

The examples in Section 2 use this interface to implement spatial and temporal discretization schemes. If each time step of a PDE discretization is split into $N$ sub-timesteps, then an application developer would implement a series of sub-timestep as functions and calls those properly in our timestepping interface function.

This interface, defined by the ``SpatialPoint" and the ``Global Point or Sub-timestep" indices, separates the concerns of application developers and computer scientists. A numerical analysts can program a solver in functions that operate on individual ``SpatialPoint" structures. His code would be lightweight, thus easier to verify and to validate. It would also be portable to various current and future computer architectures. A computer scientist, on the other hand, can implement the swept rule interface to orchestrate the execution of atomic operations according to the swept rule. His code does not need to concern what these atomic operations are, thus is independent of the application. A simple working implementation can be found at \url{https://github.com/hubailmm/K-S_1D_Swept}.
\section{Swept rule for finite difference -- solution of the Kuramot-Sivashinsky equation}

The swept rule is tested on the chaotic one space dimension Kuramoto-Sivashinsky (K-S) PDE.  This equation was one of our choices to test the swept partitioning scheme as it contains high order derivatives and nonlinear terms.  K-S PDE is knows to be stiff and produce solutions that exhibit spatio-temporal chaos, as shown in Figure 5.  To solve the 1D K-S PDE, the special discretization was done using the finite difference scheme and the time integration was done using an explicit second-order Runga-Kutta integration scheme.\newline

The picture in Figure 5 shows the solution obtained to the 1D K-S PDE using swept partitioning with periodic boundary conditions and an initial condition that is given by: $u(x,0) = 2\cos\left(\frac{19x}{128}\right)$.\newline

Figure 6 shows performance comparison between the classical straight and the swept decomposition that are applied to the one space dimension K-S PDE with the same spacial discretization.  All the runs were conducted on a small two-node cluster that was formed on Amazon's Elastic Computing (EC2) services using ``StarCluster", an open source cluster-computing toolkit for Amazon's EC2\cite{starcluster}.  The EC2 instance type was ``m1.xlarge" with an Amazon Machine Image (AMI) of ``starcluster-base-ubuntu-13.04-x86\_64".\newline

As the main aim behind running this experiment is show how the swept decomposition breaks the latency barrier, the runs were conducted using a single MPI process residing in each compute node.  The CPU in each node is ``Intel(R) Xeon(R) CPU E5-2650 @ 2.00GHz" and the MPI latency between the nodes was measured and averaged to be around 150 $us$.\newline

\begin{figure}
\centering
\includegraphics[width=0.7\linewidth]{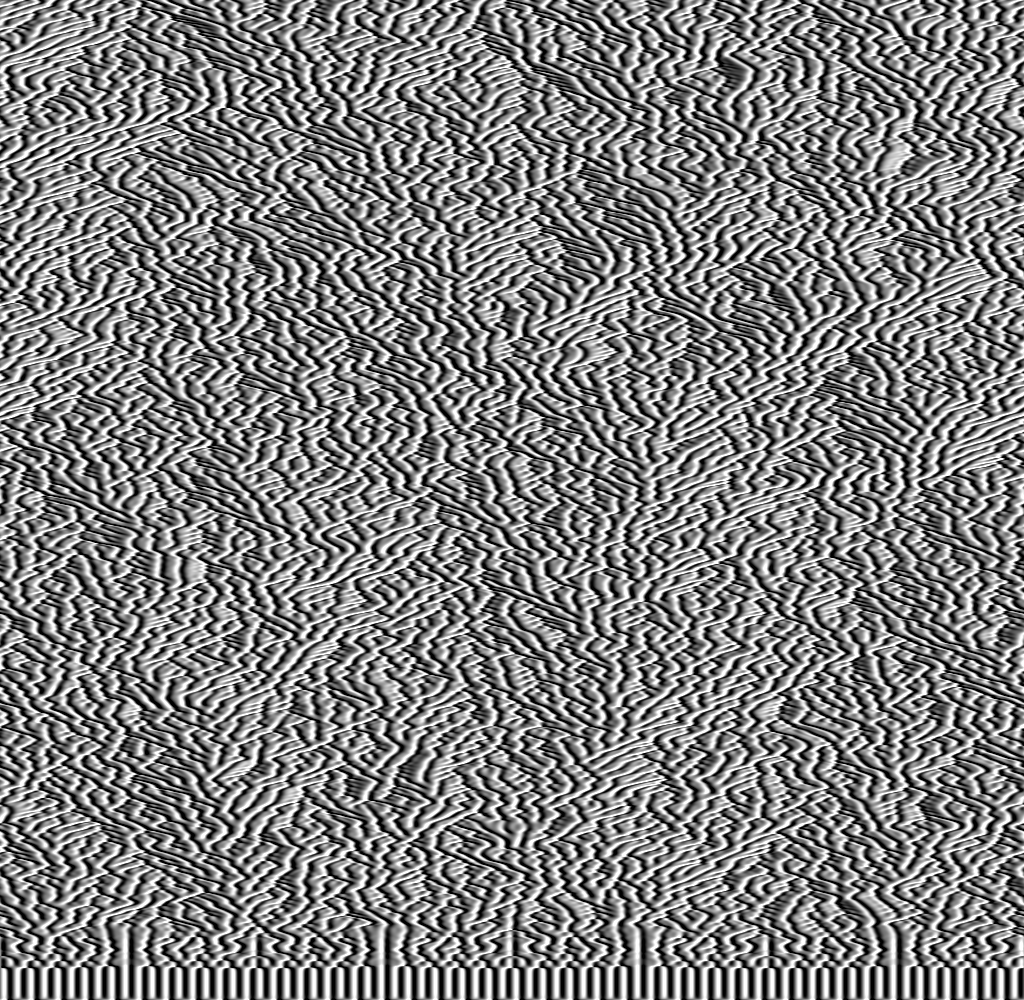}
\caption{A chaotic solution of the Kuramoto-Sivashinsky equation.  The X axis represents the space and the Y axix represents the time}
\end{figure}

Figure 6 shows that the classic domain decomposition Is limited by the latency barrier. When each node has less than 10,000 grid points, it never takes less than 150 microseconds to integrate each sub-timestep. The swept decomposition rule, in contrast, breaks this latency barrier. Whenever each node has less than 10,000 grid points, it takes less than 150 microseconds to integrate each sub-timestep. At about 100 grid points per node, it takes minimum time to integrate each time step. This optimum coincide with the intersection of the $ns$ and $2\tau/s$ lines, which is the same optimum predicted by our simplified analysis in Section 4.  At that optimum, it takes less than 10 microseconds to integrate each time step, breaking the latency barrier by a factor of over 15.\newline

A running implementation of the swept rule solving the 1D K-S PDE can found at
\url{https://github.com/hubailmm/K-S_1D_Swept}

\begin{figure}
\centering
\includegraphics[width=0.7\linewidth]{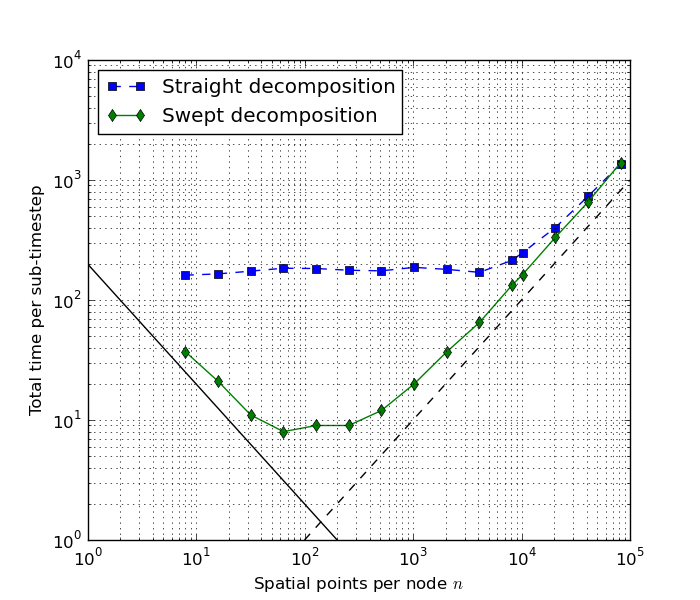}
\caption{Performance comparison between the straight and the swept decomposition when solving the 1D K-S PDE.  The black up-sloping, dashed line represent the ns term in Equation (1) , and the black down-sloping, solid lines represent the $\tau/n$.  The blue and orange lines represent the reported time per sub-timestep values for the straight and swept decompositions accordingly}
\end{figure}

\section{Swept rule for finite volume -- solution of the Euler equations of gas dynamics}

The swept rule is tested on the 1D Euler equation for gas dynamics. We discretized the Euler equation with a second order finite volume scheme. Interface flux is computed with minmod limiter and scalar dissipation proportional to the spectral radius of the Roe matrix. A second order Runge Kutta scheme, also known as the midpoint rule, is used for time integration. The same discretization of the Euler equations is implemented both in classic and swept decomposition rules; their source code can be found at
\url{https://github.com/qiqi/Swept1D/blob/master/euler_classic.cpp} and \url{https://github.com/qiqi/Swept1D/blob/master/euler_swept.cpp} .

\begin{figure}
\centering
\includegraphics[width=0.6\linewidth]{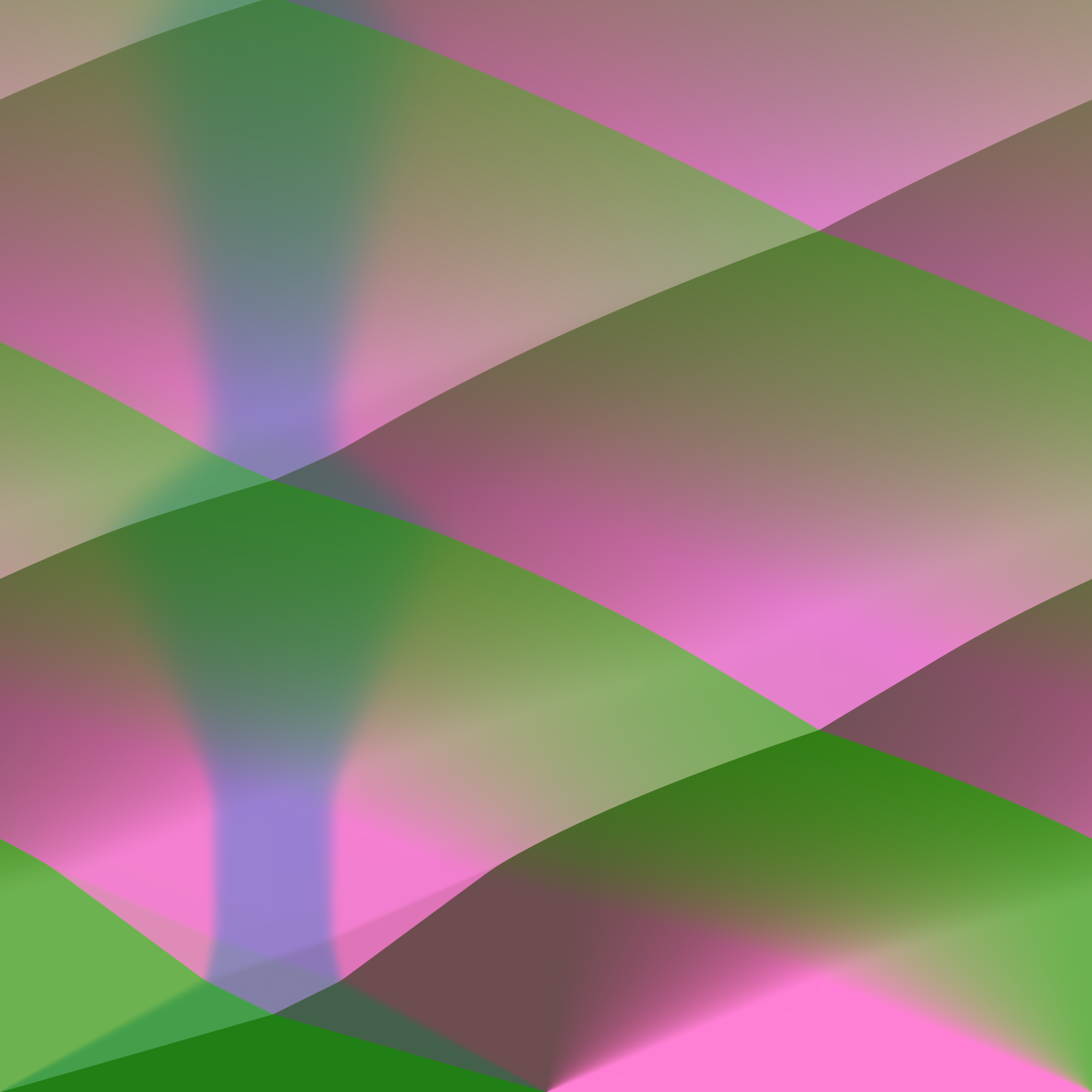}
\caption{The solution to Euler's PDE obtained using the swept rule decomposition.  The X axis represents the space and the Y axix represents the time}
\end{figure}

The picture in Figure 7 shows our solution to the Euler equation. We use red, green and blue to visualize densities of mass, momentum and energy, using PngWriter by Frank Ham\cite{png}. The spatial domain is periodic, and solution is initialized according to the SOD shock tube test case. The initial condition is $\rho=0.125, u=0, p = 0.1$ on the left half of the domain, and $\rho=1, u=0, p = 1$ on the right half of the domain.\newline

Figure 8 summarizes the performance of the straight and swept decomposition rules, applied to the same discretization of the Euler equation. Both straight and swept decomposition are implemented with a flat MPI architecture. The performance is measured on 64 MPI processes running on 8 nodes on the voyager cluster at MIT, each node containing an Intel Xeon Processor E5-1620 quad core processor with hyperthreading.\newline

Figure 8 shows that the classic domain decomposition Is limited by the latency barrier. When each node has less than 1,000 grid points, it never takes less than 60 microseconds to integrate each sub-timestep. The swept decomposition rule, in contrast, breaks this latency barrier. Whenever each node has less than 1000 grid points, it takes less than 60 microseconds to integrate each sub-timestep. At about 50 grid points per node, it takes minimum time to integrate each time step. This optimum coincide with the intersection of the $ns$ and $2\tau/s$ lines, which is the same optimum predicted by our simplified analysis in Section 4.  At that optimum, it takes less than 6 microseconds to integrate each time step, breaking the latency barrier by a factor of about 10.

\begin{figure}
\centering
\includegraphics[width=0.8\linewidth]{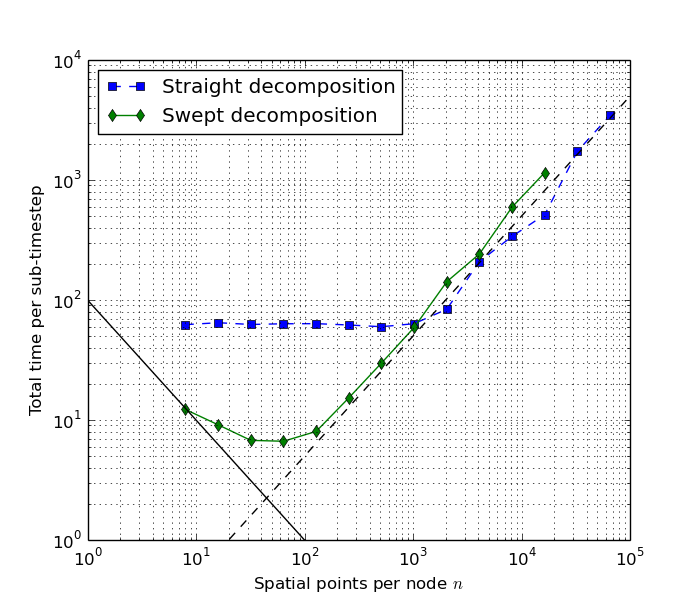}
\caption{The performance of the straight and swept decomposition rules to Euler's equation.  The black up-sloping, dashed line represent the ns term in Equation (1) , and the black down-sloping, solid lines represent the $\tau/n$.  The blue and green lines represent the reported time per sub-timestep values for the straight and swept decompositions accordingly}
\end{figure}
\section{Conclusion}

We introduced the swept rule for decomposing space and time in solving PDEs. The swept rule breaks the latency barrier, advancing each sub-timestep in a fraction of the time it takes for a message to travel from one computing node to another. In our experiment with the Kuramoto-Sivashinsky equation on Amazon EC2, over 15  sub-timesteps can be integrated during each latency time. In another experiment with the Euler equation on a Ethernet-based cluster, about 10 sub-timesteps can be integrated during each latency time.\newline

To examine how fast the swept space-time decomposition rule can be, we conducted a simplified theoretical analysis. The analysis shows that time integration can be made faster not only by reducing latency, but also by beefing up each node. 1 nanosecond per time step may be possible with currently commissioned hardware, but only if the the power of each node can be fully utilized.\newline

The swept decomposition rule can be implemented through an interface, which separates numerical schemes from its computational implementation. The interface allows different numerical schemes to use the same implementation of the swept decomposition rule. It is also easy to switch between different implementations sharing the same interface. The authors' implementation of the swept decomposition scheme, along with testcases and illustrations in this article, are open source.\newline

\subsection*{Acknowledgment}
We acknowledge the Advanced Research Center at Saudi Aramco for sponsoring graduate students to pursue research in the Computational Science and Engineering area.  We also acknowledge NASA NRA Award 15-TTT1-0057 under Dr. Eric Nielsen and Dr. Mujeeb Malik, AFOSR Award F11B-T06-0007 under Dr. Fariba Fahroo, DOE award DE-FG02-14ER26173/DE-SC00011089 under Dr. Sandy Landsberg and NASA Award NNH11ZEA001N under Dr. Harold Atkins.

\section*{References}

\bibliography{mybibfile}

\end{document}